\documentclass[conference]{IEEEtran}
\IEEEoverridecommandlockouts
\usepackage{cite}
\usepackage{amsmath,amssymb,amsfonts}
\usepackage{algorithmic}
\usepackage{graphicx}
\usepackage{textcomp}
\usepackage{tikz}
\usepackage{tikzsymbols} 
\usetikzlibrary{calc,arrows}
\usetikzlibrary{chains}
\usetikzlibrary{shapes.geometric}
\usepackage{graphicx}
\usepackage{color}
\usepackage{booktabs}
\usepackage{textcomp}

\usepackage{enumitem}

\def\BibTeX{{\rm B\kern-.05em{\sc i\kern-.025em b}\kern-.08em
    T\kern-.1667em\lower.7ex\hbox{E}\kern-.125emX}}
\begin{document}

\newcommand{\codeft}[1]{\mbox{\fontfamily{qcr}\selectfont #1}}
\newcommand{\tab}[1][0.6cm]{\hspace*{#1}}

\title{\codeft{ferify}: A Virtual Machine File Protection System against Zero-Day Attacks \\ [0.1in]
{\large December 2017}
}


\author{\IEEEauthorblockN{Alexis Peppas}
\IEEEauthorblockA{\textit{Helenic Navy}\\
Greece \\
apeppas@outlook.com}
\and
\IEEEauthorblockN{Geoffrey G. Xie}
\IEEEauthorblockA{\textit{Computer Science Department} \\
\textit{Naval Postgraduate School}\\
xie@nps.edu}
\and
\IEEEauthorblockN{Charles D. Prince}
\IEEEauthorblockA{\textit{Computer Science Department} \\
\textit{Naval Postgraduate School}\\
cdprince@nps.edu}
}



\maketitle

\begin{abstract}

Most existing solutions for protecting VMs assume known attack patterns or signatures and focus on detecting malicious manipulations of system files and kernel level memory structures. In this research we develop a system called \codeft{ferify}, which leverages VM introspection (VMI) to protect \emph{user files} hosted on a VM against unauthorized access even after an attacker has managed to obtain root privileges on the VM. \codeft{ferify} maintains in the hypervisor domain a shadow file access control list (SACL) that is totally transparent to the VM. It uses the SACL to perform independent access control on \emph{all} system calls that may operate on the target files.  Further, \codeft{ferify} prevents kernel modification, ensures the integrity of process ownership, and supports hypervisor based user authentication. We have developed a \codeft{ferify} prototype for Linux and through a set of controlled experiments we show that the system is able to mitigate a range of zero-day attacks that otherwise may evade signature-based solutions.  In addition, we analyze the root cause of the observed high processing overhead from trapping of system calls, and propose a general solution that can potentially cut that overhead by half.

\end{abstract}


\section{Introduction}\label{sec:intro}

A successful attack on a computer system typically results in the attacker obtaining the \emph{root} privilege for the system. If the computer system is a virtual machine (VM), this means that the attacker has total access to all files hosted on the VM. How to detect and contain this type of root-kit attacks remains an important security problem.

While virtualization brings about new security challenges specific to VM operation, it also offers new solution approaches. The research community has long recognized the unique vantage point provided by the hypervisor for VM monitoring and malware detection. In particular, the VM introspection (VMI) capabilities~\cite{b15,b11} have shown great promise. We observe two main advantages by deploying security solutions on the hypervisor. First, the code base is relatively small due to its narrow focus and thus, is relatively easy to catch software bugs or presence of malware. Second, being part of the critical path for accessing physical resources, the hypervisor is able to exert \emph{independent} and process/thread level control over program executions on a guest VM. And this control can be dynamic, e.g., revoking a user's permission to access certain resources without rebooting the VM.


\par Paladin~\cite{b1} 
is among the first systems that leverage VMI 
to detect malicious manipulations of system files and/or run-time data structures and contain such attacks by aborting the offending processes and rolling back suspicious data modifications. However, this system requires a trusted software helper module running in the guest VM~\cite{b1}. More importantly, Paladin as well as most other existing solutions for protecting VMs focus on protecting systems files and other data against known attack signatures. Thus, these solutions may have limited power against zero-day attacks.

%
%

\par In this paper, we present the design and evaluation of a \emph{user centric} VM file protection system which we call~\codeft{ferify}. As an overarching goal, we seek to protect selected user files hosted on a VM against unauthorized access \emph{in spite of} a successful attack on the VM.  \codeft{ferify} is implemented as a DRAKVUF~\cite{b11} 
plugin that is completely independent from the guest VM. It maintains a separate file access control list (ACL) in the hypervisor and performs access control on all system calls that  may operate on the target files. Additionally, we constrain kernel modifications and ensure the integrity of process ownership and other critical data in order to mitigate a range of zero-day attacks.

The rest of the paper is organized as follows. We review related work and provide background information in Section~\ref{sec:related}. We present the design and implementation details of \codeft{ferify} in Sections~\ref{sec:design} and~\ref{sec:impl}, respectively. A detailed evaluation of the system's file protection capabilities and its processing overhead is provided in Section~\ref{sec:eval}, followed by a detailed analysis of the overhead along with a solution for reducing it in Section~\ref{sec:overhead}. Finally, we discuss possible extensions and its current limitations in Section~\ref{sec:discussion} and then offer some concluding remarks in Section~\ref{sec:conclude}.

\section{Related Work}\label{sec:related}

\begin{table*}[t]
	\footnotesize
	\centering
	\caption{Overview of Existing solutions.}
	\label{tbl:overview}
	\begin{tabular}{l|cccccc}
		\toprule
		&  &  & \multicolumn{2}{c}{\textbf{Kernel Protection}}  & \multicolumn{2}{c}{\textbf{File Protection}}  \\
		\textbf{Solution}& \textbf{In-VM} & \textbf{Out-VM} & \scriptsize {\textbf{Detection}} &  \scriptsize {\textbf{Prevention}} & \scriptsize {\textbf{Detection}} & \scriptsize {\textbf{Prevention}} \\

		\hline

		Virtuoso~\cite{b6},~SIM~\cite{b19} 					& \checkmark & - & \checkmark & - & - & -\\

		\hline

		Haven~\cite{b2},~Overshadow~\cite{b4},					& & & & & & \\
		InkTag~\cite{b9},~Lares~\cite{b15},~SHype~\cite{b17} 					& \checkmark & - & - & \checkmark & - & -\\

		\hline
		
		Crawford and Peterson~\cite{b5},~ReVirt~\cite{b7},					& & & & & & \\
		VMI~\cite{b8},~VMWatcher~\cite{b10},					& - & \checkmark & \checkmark & - & - & -\\
		Macko et al.~\cite{b12},~PoKeR~\cite{b16},					&  &  &  &  &  & \\
		Strider, Ghostbuster~\cite{b22}					&  &  &  &  &  & \\

		\hline

		SecVisor~\cite{b18},~Srinivasan et al.~\cite{b20}					& - & \checkmark & - & \checkmark & - & -\\
		Sentry~\cite{b21},~HUKO~\cite{b23}					&  &  &  &  & & \\

		\hline

		Nasab~\cite{b13}	& - & \checkmark & - & \checkmark & \checkmark & -\\

		\hline

		Paladin~\cite{b1}					& \checkmark & - & - & \checkmark & \checkmark & \checkmark\\

		\bottomrule
	\end{tabular}	
	\label{prior-solutions}
\end{table*}

In this section, we first review existing security solutions that are most related to this work and then provide a short description of the DRAKVUF and LibVMI software, upon which \codeft{ferify} has been built.

{\bf VM monitoring and protection:} 
Table~\ref{prior-solutions} presents a summary of seventeen relevant solutions found in the literature. Some are integrated into the VM (i.e., ``in-VM'') while others leverage the hypervisor (i.e.,``Out-VM''). Most focus on ensuring kernel integrity and protecting kernel level data structures such as the process control block.  Only two solutions~\cite{b1,b13}
provide some level of file protection; however, they require known attack signatures.

{\bf DRAKVUF / LibVMI:}  LibVMI~\cite{b15-1} is a C library developed by the Sandia Labs to simplify the development of VM introspection solutions for the Xen hypervisor. DRAKVUF~\cite{b11} is a malware analysis tool built upon LibVMI. It supports in-depth execution tracing of arbitrary binaries (including kernel processes) that is totally transparent to the VM being monitored. Furthermore, it is designed to be extensible by supporting plugins.

\section{Design}\label{sec:design}

In this section, we first describe the threat model and high level requirements and then present the details of how \codeft{ferify} traps system calls and is able to continue file access control even after the root account of the target VM has been compromised. 

\subsection{Threat Model and Requirements}

We have designed \codeft{ferify} by assuming this threat model:
\begin{itemize}[leftmargin=0.14in]
  \item  The hypervisor is considered secure; how to protect the hypervisor is outside the scope of this research. 
  \item The protected files are only remotely accessible and protected by a public-key authentication and encryption scheme (i.e., \emph{SSH}).
  \item The private keys of authorized users are secure, while the attacker may have obtained {root} privileges on the VM through a successful attack.
\end{itemize}

Additionally, we have set these high level design objectives:

\begin{itemize}[leftmargin=0.14in]
	\item The solution must be \emph{completely} out-VM to avoid subversion from the VM side. 
	\item The protected VM must remain usable by authorized users. In particular, the extra processing overhead incurred by the solution must be bounded to a tolerable range.
\end{itemize}

\subsection{Basic Functionality}

\begin{figure}[!htb]
\centering
\begin{tikzpicture}
\small
\node[anchor= north west, fill=red!20, rectangle, rounded corners, draw=black, minimum width=30mm, minimum height=5mm, label={[yshift=-0.55cm]north:Vulnerable VM}] at (0, 0) {};

\node[anchor= north west, fill=blue!30, rectangle, rounded corners, draw=black, minimum width=30mm, minimum height=45mm, label={[yshift=-0.5cm]north:Protection Zone}] at (4, -0.5) {};

\node[anchor= north west, fill=blue!20, rectangle, rounded corners, draw=black, minimum width=20mm, minimum height=8mm, label={[yshift=-0.8cm, align=center, font=\scriptsize]north:DRAKVUF /\\LibVMI}] at (0.5, -1.25) {};

\node[anchor= north west, fill=blue!20, rectangle, rounded corners, draw=black, minimum width=25mm, minimum height=8mm, label={[yshift=-0.65cm, align=center, font=\footnotesize]north:\codeft{ferify} plugin}] at (4.25, -1.5) {};

\draw[anchor= south west, draw=red!70, line width=0.5mm,->] (2.5, 0.5) -- (2.5, 0);    
\node[anchor= north west, minimum width=5mm, minimum height=7mm,  label={[yshift=0cm, xshift=-0.25cm, align=center, font=\scriptsize ]north:Attacker}] at (2.5, 0.5) {};

\node[anchor= north west, minimum width=20mm, minimum height=8mm, label={[yshift=-0.7cm, align=center, font=\scriptsize]north:File or kernel operation}] at (-0.5, -0.5) {};

\draw[anchor= south west, draw=blue!70, line width=0.5mm,->] (0.5, 0.5) -- (0.5, 0);
\node[anchor= north west, minimum width=5mm, minimum height=7mm,  label={[yshift=0cm, xshift=-0.25cm, align=center, font=\scriptsize ]north:Authorized user}] at (0.5, 0.5) {};

\draw[anchor= south west, draw=black, line width=0.3mm,->] (1.5, -0.5) -- (1.5, -1.25);
\draw[anchor= south west, draw=black, line width=0.3mm,->] (2.5, -1.75) -- (4.25, -1.75);
\node[anchor= north west, minimum width=20mm, minimum height=8mm, label={[yshift=-0.3cm, xshift=0.5cm, align=center, font=\scriptsize]north:Trapped\\ sys-calls}] at (1.6, -1.5) {};


\node[anchor= north west, minimum width=20mm, minimum height=8mm, label={[yshift=-0.35cm, align=center, font=\scriptsize]north:No}] at (-0.25, -2.5) {};

\node[regular polygon, regular polygon sides=8, minimum width=1cm, draw, line width=1pt](n1) at (-0.1, -2.8){};
\node[regular polygon, regular polygon sides=8, minimum width=1cm, draw=white,fill=red!80,label=center:\color{white}{\scriptsize\bfseries\sffamily STOP}]at(n1) at (-0.1, -2.8){};

\node[anchor= north west, minimum width=20mm, minimum height=8mm, label={[yshift=-0.35cm, align=center, font=\scriptsize]north:Yes}] at (0.25, -3.3) {};
\node[inner sep=0pt] (russell) at (1.75,-3.7)
    {\includegraphics[width=.4cm]{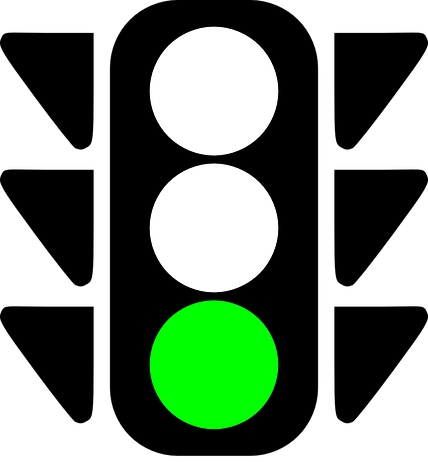}};


\draw[anchor= south west, draw=black, line width=0.3mm,->] (5, -2.3) -- (5, -2.6);

\draw[anchor= south west, draw=black, line width=0.2mm,-] (5, -2.6) -- (5.45, -2.9) -- (5, -3.2) -- (4.55, -2.9) -- (5, -2.6);
\draw[anchor= south west, draw=black, line width=0.3mm,->] (5, -3.2) -- (1.5, -3.2) -- (1.5, -4);
\draw[anchor= south west, draw=black, line width=0.3mm,->] (4.55, -2.9) -- (0.38, -2.9);


\node[anchor= north west, fill=white, rectangle, rounded corners, draw=black, minimum width=30mm, minimum height=30mm, label={[yshift=-0.5cm]north:HDD}] at (0, -4) {};
  
\node[anchor= north west, fill=white, rectangle, draw=black, minimum width=5mm, minimum height=7mm, label={[yshift=-0.5cm]north:}] at (0.2, -4.8) {};
\node[anchor= north west, fill=white, rectangle, draw=black, minimum width=5mm, minimum height=7mm, label={[yshift=-0.5cm]north:}] at (0.3, -4.9) {};
\node[anchor= north west, fill=white, rectangle, draw=black, minimum width=5mm, minimum height=7mm, label={[yshift=-0.5cm]north:}] at (0.4, -5) {};
\node[anchor= north west, fill=white, rectangle, draw=black, minimum width=5mm, minimum height=7mm,  label={[yshift=-1.5cm, xshift=-0.15cm, align=center, font=\scriptsize ]north:Other\\files}] at (0.5, -5.1) {};

\node[anchor= north west, fill=blue!30, rectangle, draw=black, minimum width=5mm, minimum height=7mm, label={[yshift=-0.5cm]north:}] at (2, -4.8) {};
\node[anchor= north west, fill=blue!30, rectangle, draw=black, minimum width=5mm, minimum height=7mm, label={[yshift=-0.5cm]north:}] at (2.1, -4.9) {};
\node[anchor= north west, fill=blue!30, rectangle, draw=black, minimum width=5mm, minimum height=7mm, label={[yshift=-0.5cm]north:}] at (2.2, -5) {};
\node[anchor= north west, fill=blue!30, rectangle, draw=black, minimum width=5mm, minimum height=7mm,  label={[yshift=-1.5cm, xshift=-0.15cm, align=center, font=\scriptsize ]north:Protected\\files}] at (2.3, -5.1) {};

\draw[anchor= south west, draw=black, line width=0.3mm,->] (6, -2.3) -- (6, -3.7);

\draw[fill=blue!20] (6, -4.1) ellipse (0.3cm and 0.1cm);
\draw[fill=blue!20] (6, -4) ellipse (0.3cm and 0.1cm);
\draw[fill=blue!20] (6, -3.9) ellipse (0.3cm and 0.1cm);
\node[anchor= north west, minimum width=7mm, minimum height=1mm , label={[yshift=-1cm, xshift=0.05cm, align=center, font=\small]north:SACL}] at (5.6, -3.7) {};
\draw[fill=blue!20] (6, -3.8) ellipse (0.3cm and 0.1cm);

\draw[anchor= south west, draw=black, line width=0.1mm,-] (5.7, -3.8) -- (5.7, -4.1);
\draw[anchor= south west, draw=black, line width=0.1mm,-] (6.3, -3.8) -- (6.3, -4.1);

\end{tikzpicture}\caption{Design overview}
\label{fig:design}
\end{figure}
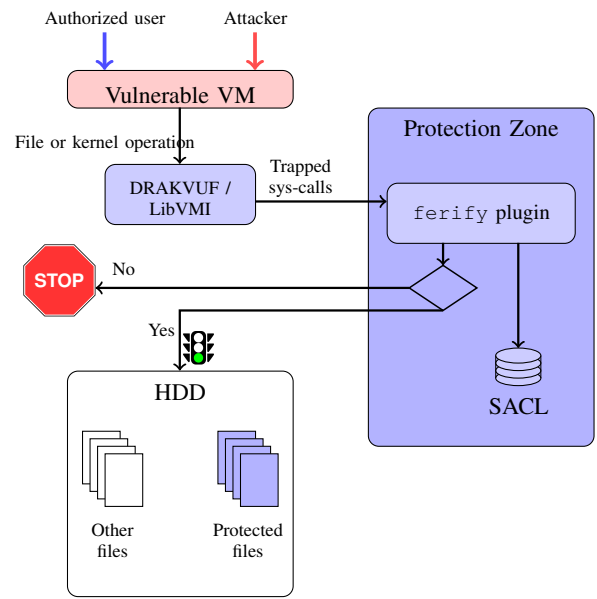

The high level design of \codeft{ferify} for Linux and Xen is illustrated in Fig.~\ref{fig:design}. In a nutshell, the system traps all relevant system calls and uses a shadow access control (SACL) maintained inside the hypervisor to perform file access control.  \codeft{ferify} identifies authorized user accounts by their \codeft{uid} and \codeft{gid} in the SACL; therefore, it includes additional security measures to ensure the integrity of (i) user and group identification, (ii) process ownership, and (iii) the kernel, as well as two further extensions to enhance security, which will be presented in the succeeding sub-section \emph{C}.

{\bf Trapping of System Calls:}~
We use DRAKVUF~\cite{b11} to trap a specific set of system calls that are relevant to file operations and kernel access, as listed in Table \ref{tbl:syscalls}. A trap in our case is a software breakpoint (opcode \codeft{0xCC} or the so-called \codeft{INT 3} instruction for an Intel x86 CPU) injected at the beginning of all trapped system calls. The behavior can be compared to that of a debugger. LibVMI allows for the assignment of a callback function that gets executed when a trap gets caught. \codeft{ferify} provides a callback function that checks the validity of each trapped system call. 

\begin{table}[ht]
	\centering
	\caption{Trapped system calls}
	\label{tbl:syscalls}
	\begin{tabular}{cc}
		\footnotesize{\fontfamily{qcr}\selectfont open()} 					&
		\footnotesize{\fontfamily{qcr}\selectfont openat()} 				\\
		\footnotesize{\fontfamily{qcr}\selectfont name\_to\_handle\_at()*} 	&
		\footnotesize{\fontfamily{qcr}\selectfont open\_by\_handle\_at()*} 	\\
		\footnotesize{\fontfamily{qcr}\selectfont rename()} 				&
		\footnotesize{\fontfamily{qcr}\selectfont renameat()} 				\\
		\footnotesize{\fontfamily{qcr}\selectfont renameat2()} 		    	&
		\footnotesize{\fontfamily{qcr}\selectfont unlink()} 			    \\	
		\footnotesize{\fontfamily{qcr}\selectfont unlinkat()}   			&	
		\footnotesize{\fontfamily{qcr}\selectfont truncate()} 	    		\\
		\footnotesize{\fontfamily{qcr}\selectfont link()} 					&
		\footnotesize{\fontfamily{qcr}\selectfont linkat()} 				\\
		\footnotesize{\fontfamily{qcr}\selectfont symlink()} 				&
		\footnotesize{\fontfamily{qcr}\selectfont symlinkat()} 				\\
		\footnotesize{\fontfamily{qcr}\selectfont execve()} 				&
		\footnotesize{\fontfamily{qcr}\selectfont execveat()} 				\\
		\footnotesize{\fontfamily{qcr}\selectfont exit()}   				&
		\footnotesize{\fontfamily{qcr}\selectfont exit\_group()} 			\\
		\footnotesize{\fontfamily{qcr}\selectfont init\_module()*} 			&
		\footnotesize{\fontfamily{qcr}\selectfont finit\_module()*} 		\\
		\footnotesize{\fontfamily{qcr}\selectfont kexec\_load()*} 			&
		\footnotesize{\fontfamily{qcr}\selectfont } 				        \\
		
	\end{tabular}	
\end{table}

Specifically, the callback function retrieves the arguments of the system call from the CPU registers according to the 64-bit Linux system call convention and then performs all the necessary checks against the SACL. The format of the SACL is shown in Fig.~\ref{fig:sacl}. The ``Permission'' field is set following the Linux three-octet file permission convention (\codeft{u}$\mid $\codeft{g}$\mid $\codeft{o}), referring to the permission for the user owning the file, users of the same group, and all other users, respectively. The three bits of each octet represent the read, write, and execute permission flag, respectively. Therefore, for the example in Fig.~\ref{fig:sacl}, \codeft{644} means that the owner of the file can read and write to the file while all other users including those of the same group have only read access. \codeft{400} means that the owner of the file can read the file, while no other users have access to the file. The ``User'' and ``Group'' fields provide the \codeft{uid} and \codeft{gid} of the file owner, respectively, with a special value of \codeft{0} referring to the root user or the root user group.
Finally, if the file in question does not have an entry in the SACL, \codeft{ferify} deems it noncritical and the check successful.

\begin{figure}[ht]
    \scriptsize
	\centering
	\begin{tabular}{lccc}
Full path file or directory name&Permission &User&Group\\
\hline
\codeft{/home/user/Documents/critical.txt} & 
	\codeft{644} & \codeft{1000} & \codeft{1000}\\
\codeft{/home/user/Desktop/read-only.pdf}	 & 
	\codeft{400} & \codeft{1000} & \codeft{1000}\\
\codeft{/etc/shadow}			 & 
	\codeft{220} & \codeft{0}    & \codeft{0}\\
	\hline
	\end{tabular}
	\caption{Example SACL entries.}
	\label{fig:sacl}
\end{figure}

The system call is allowed to proceed \emph{only if} it passes the SACL check.  When the decision is to deny the system call, \codeft{ferify} replaces the pointer that holds the file-name with a \codeft{NULL} pointer; doing so allows the system call to proceed but eventually fail when it tries to de-reference a NULL pointer.

We have successfully trapped and processed all system calls listed in Table~\ref{tbl:syscalls}, albeit we currently simply deny those system calls marked with an asterisk due to time constraints for carrying out the implementation. 

%
{\bf User \& Group Integrity:}~
\codeft{ferify} prohibits switching of user accounts through the \codeft{su} command, by denying all users including the root the write permission to the \codeft{/etc/pam.d/su} file. Similarly, it enforces no write to password files \codeft{/etc/passwd} and \codeft{/etc/shadow} by adding deny entries in the SACL. This may raise usability challenges for authorized users. However, we envision that the VM will be periodically taken offline for maintenance and in these offline periods, these account restrictions can be lifted.

%
{\bf Process Ownership Integrity:}~
For each running user process, \codeft{ferify} keeps track of the \codeft{uid} and \codeft{gid} of its creator. This tracking allows for monitoring of malicious attempts to change this ownership information. Specifically,  \codeft{ferify} traps the \codeft{ret\_from\_fork} kernel symbol and store in the hypervisor a hashtable of the owner information for all legitimate processes created by authorized users. (It also traps \codeft{clone()} to monitor all new processes.)
 This hashtable is then used to determine if a trapped system call is indeed made by an authorized user\footnote{If there is a legitimate change in the ownership of a process (through \codeft{sudo}) we update the stored information. This update is important in order to retain system \emph{usability}.} before the SACL check. 

%
{\bf Kernel Integrity:}~
By default, a (malicious) root user can modify kernel structure data and load new kernel modules. To mitigate this attack vector, \codeft{ferify} traps and blocks the \codeft{init\_mod()} and \codeft{finit\_mod()} system calls. It also write-protects \codeft{/etc/modules}, and possibly additional files and folders depending on the Linux distribution, to prevent the loading of new kernel modules during boot time. Doing so clearly imposes some usability issues. As discussed earlier, we envision to perform the legitimate kernel modifications in specific maintenance periods when the VM is taken offline and after properly authenticating and checking of the integrity of the new kernel modules.  


To protect against malicious kernel swapping, \codeft{ferify} blocks the system call \codeft{kexec\_load()}, which loads a new kernel for later execution. This introduces a usability limitation, by not allowing automatic kernel updates, some of which are necessary to fix kernel bugs. Since this operation can be performed in a more controlled environment (i.e., during offline periods) at the administrator's discretion, we expect this to be a reasonable limitation. 



\subsection{Two Extensions}

To further enhance security, we extend \codeft{ferify} in two aspects.

%
{\bf 2-Step Authentication:}~
We have integrated into \codeft{ferify} a two-step authentication mechanism leveraging the hypervisor as a user authenticating agent. The authentication is based on a pre-configured shared secret between each authorized user and the hypervisor, and performed as part of the callback function for the \codeft{open()} system call. Specifically, when this option is turned on for an authorized user, even after the user has successfully logged into the VM, \codeft{ferify} still considers the user ``unauthenticated'' until he/she passes a second form of authentication as follows. The user must prove that he/she possesses the shared secret by presenting a challenge response pair of strings to \codeft{ferify}, the latter of which is the SHA512 hash of the former concatenated with the shared secret, through an \codeft{open()} system call. For example, the user may trigger this authentication through the \codeft{touch} command and encode the challenge response strings in the file-name argument along with an artificial path-name to avoid collision with a real file. 


After \codeft{ferify} verifies that the challenge response strings are valid, it changes the user's status to ``authenticated'' for a predetermined time-frame, during which the user's file access permission will be according to the SACL; otherwise, the user remains in the ``unauthenticated'' status and will not be granted access to any file specified in the SACL.

%
{\bf Program Execution White-Listing:}~ 
We have added an option to the callback functions of the \codeft{execve()} and \codeft{execve\_at()} system calls for \codeft{ferify} to deny, instead of permit by default, execution of a file that is not listed in the SACL. In other words, when this option is turned on, a file can be executed only if it has a  permit entry in the SACL. Effectively, this option makes the SACL a white-list for program execution. Again, there is a trade-off of usability with this option, but we believe this option is useful for certain deployment scenarios.

\section{Implementation}\label{sec:impl}

We have created a prototype implementation of \codeft{ferify} for Linux, as a plugin for the v0.5-655884f version of DRAKVUF, which is bundled with the 0.12 release of LibVMI and the 4.8.1 version of Xen. The implementation consists of about 2,400 lines of C code\footnote{We will make all source code publicly available once the double blind requirement for this paper is lifted.}. The SACL is implemented as a hashtable to bound the search processing overhead. To assist testing with different SACL sizes, we have also created a script that retrieves information for all files currently residing in a given VM and creates an artificial \emph{full} SACL with a ``permit'' entry for each of the files.

We have chosen a computer with an Intel i7-6700 CPU as the test platform, given that DRAKVUF~\cite{b11} is designed to take advantage of hardware virtualization extensions found in Intel CPUs. 
The computer is equipped with 8 GB of RAM. It runs the Ubuntu 16.04 64-bit version of Linux, more specifically, the 4.10.0-30-generic kernel, as the host operating system (OS) for Dom0. The guest VM also runs Ubuntu 16.04 64-bit, but with the 4.8.0-54-generic kernel. 

\section{Evaluation}\label{sec:eval}

Our evaluation of \codeft{ferify} consists of two parts. First, we validate its ability to protect files against unauthorized access, particularly its potential to mitigate zero-day attacks. Second, as \codeft{ferify} needs to trap more than a dozen system calls, we quantify expected performance degradation upon authorized users due to the extra processing overhead it introduces. 


\subsection{Validation of File Protection}

Ideally, we can enumerate the exact range of attacks that \codeft{ferify} is able to mitigate and perform experiments to confirm the effectiveness. However, given the unpredictable nature of zero-day attacks, and the amount of effort required to hypothesize and enact the likely large number of attacks to ensure coverage, we take an alternative, more practical, approach, whereby we show the main design features as presented in Section~\ref{sec:design} achieve their objectives.  More specifically, we consider two types of access to protected files hosted on the guest VM: one by an authorized user and the other by an attacker, as illustrated in Fig.~\ref{fig:design}. In line with our threat model, we assume that (i) the authorized user's private key is secure and 
(ii) the attacker has gained root privileges on the guest VM through 
a compromise originating from a different user ID than the legitimate user. 

To validate \codeft{ferify}'s basic functionality, we performed specific file operations to emulate the actions of the attacker. We try these operations with different permissions defined in the SACL each time, to verify that the permissions we set are actually enforced. The SACL we used contains all files of the VM system.\footnote{The VM was a fresh installation of Linux with only a few essential packages added, such as gcc, rekall, and openssh-server. The SACL for this setup contained more than 400,000 entries.} The results of the tests show that we can successfully deny unauthorized access to protected files, even if the OS of the VM would allow it given that the attacker has gained root privileges. 

{\bf User \& Group Integrity:}~
As our tests reveal, we can limit access to files on a per user basis. This constraint applies equally to all users, including the root user. Fig.~\ref{fig:root_sacl} shows the SACL implementation for the \codeft{/etc/shadow} and \codeft{/etc/pam.d/su} files, and Fig.~\ref{fig:passwd_deny} shows the result of the root user trying to change another user's password, after the \codeft{/etc/shadow} file has been write protected in the SACL.

\begin{figure}[ht]
\footnotesize
\centering
\begin{tabular}{lccc}
	Full Path File Name&Permissions&User&Group	\\
	\hline
	\codeft{/etc/shadow}  &  \codeft{400} & \codeft{0} & \codeft{0}\\
	\codeft{/etc/pam.d/su} & \codeft{000} & \codeft{0} & \codeft{0}\\
	\hline
\end{tabular}
	\caption{SACL entries preventing root user from overwriting two system files.}
	\label{fig:root_sacl}
\end{figure}

\begin{figure}[ht]
	\footnotesize{\fontfamily{qcr}\selectfont 
		root@HVM-domU:\~\# passwd alice \\
		Enter new UNIX password: \\
		Retype new UNIX password: \\
		passwd: Authentication token manipulation \textcolor{red}{error} \\
		passwd: password \textcolor{red}{unchanged} \\
		root@HVM-domU:\~\# }
	\caption{Confirmation of denial of password change from \emph{root} account.}
	\label{fig:passwd_deny}
\end{figure}

In addition, we validated that \codeft{ferify} is able to prevent escalation of user's privilege to root through the \codeft{sudo} command, as illustrated by   
Fig.~\ref{fig:sudo_deny}. For brevity we omit the actual SACL entries for implementing this policy. 

\begin{figure}[ht]
	\footnotesize{\fontfamily{qcr}\selectfont 
		user@HVM-domU:\~\$ sudo ls \\
		sudo: unable to open /etc/sudoers: \textcolor{red}{Bad address} \\
		sudo: \textcolor{red}{no} valid sudoers sources found, quitting  \\
		sudo: \textcolor{red}{unable} to initialize policy plugin \\
		user@HVM-domU:\~\$ }
	\caption{Confirmation of failure of \emph{sudo}.}
	\label{fig:sudo_deny}
\end{figure}

From these experimental results, we extrapolate that \codeft{ferify} can enforce some files to be \emph{immutable}, by removing the write permissions from all users, including the root.\footnote{Paladin~\cite{b1} can also achieve this effect.} Additionally, by removing only the read permissions on system log files, we allow for normal logging operation, but an attacker cannot read the log-file and select which entries to delete, to hide any malicious activity, making the ``hiding of tracks'' harder for the attacker. 

{\bf Process Ownership Integrity:}~
From the literature, to the best of our knowledge, the only effective way to alter a process's ownership information appears to be through a kernel module. This is because that information is stored in the \codeft{task\_struct} data structure, which is part of the kernel memory space. To directly access and manipulate kernel memory variables the attacker must be able to run code as part of the kernel, i.e., using a kernel module. Therefore, the validation results for the Kernel Integrity part (which we will present next) also applies to this functionality.

{\bf Kernel Integrity:}~
Since the kernel protection mechanism consists simply of denying addition of kernel module when the VM is online, the validation was relatively straightforward. Fig.~\ref{fig:module_deny} shows the test result of trying to load a new kernel module.

\begin{figure}[ht]
	\footnotesize{\fontfamily{qcr}\selectfont 
		root@HVM-domU:\~\# insmod my\_module.ko \\
		insmod: \textcolor{red}{ERROR}: could not insert module \tab my\_module.ko: Bad file descriptor \\
		root@HVM-domU:\~\#}
	\caption{Confirmation of denying addition of new kernel module.}
	\label{fig:module_deny}
\end{figure}


{\bf 2-step Authentication:}~
In this test, we turned on the option for the authorized user. Therefore, the user was considered ``unauthenticated'' even after it had passed the user authentication by the VM and started an \codeft{ssh} session. Therefore as Fig.~\ref{fig:2step_deny} shows, the user couldn't access \codeft{test1.txt} even though the SACL contained a permit entry for the user regarding this particular file. 

\begin{figure}[ht]
	\footnotesize{\fontfamily{qcr}\selectfont 
		user@HVM-domU:\~\# echo hi > test1.txt \\
		-bash: echo \textcolor{red}{write error}: Bad file descriptor \\
 		root@HVM-domU:\~\#}
	\caption{Confirmation of failed file creation before second authentication.}
	\label{fig:2step_deny}
\end{figure}

At the next step, as Fig.~\ref{fig:2step_validate} shows, we ran the \codeft{touch} command to perform the second authentication by providing a valid pair of challenge response strings based on a shared secret pre-configured in \codeft{ferify}. It should be noted that even though the authentication was successful, an error message was outputted because of the artificial file name string. 


\begin{figure}[ht]
	\footnotesize{\fontfamily{qcr}\selectfont 
		user@HVM-domU:\~\# touch /tokens/1110d209df92a6f603f89\\
		d18e2b79dda732fe88c2fcf2347024d1b4244e1d0013723107b\\
		419e6fe7d6d0dd80b4a45d06d02271473dce873477528a67f4b\\
		b2312267 \\
		touch: \textcolor{red}{cannot touch} '/tokens/1110d209df92a6f603f89d\\
		18e2b79dda732fe88c2fcf2347024d1b4244e1d0013723107b4\\
		19e6fe7d6d0dd80b4a45d06d02271473dce873477528a67f4bb\\
		2312267': No such file or directory \\
		root@HVM-domU:\~\#}
	\caption{Illustration of second authentication using \codeft{touch}.}
	\label{fig:2step_validate}
\end{figure}

After the successful 2-step authentication, as Fig.~\ref{fig:2step_allow} shows, the user was able to create \codeft{test1.txt} as expected.

\begin{figure}[ht]
	\footnotesize{\fontfamily{qcr}\selectfont 
		user@HVM-domU:\~\# echo hi > test1.txt \\
		root@HVM-domU:\~\#}
	\caption{Confirmation of file creation after successful 2-step authentication.}
	\label{fig:2step_allow}
\end{figure}


%

{\bf Program White-Listing:}~
We turned on this option. We created a \codeft{test} program that simply prints out the string: ``Can run this program.'' We added a permit entry for this program in the SACL. We ran the program and then ran an exact copy of it, \codeft{newfile}, which was not added to the SACL. As Fig.~\ref{fig:prog_run} shows, the first execution was successful because \codeft{test} was in the white-list but the second execution failed because \codeft{newfile} was not white-listed. 

\begin{figure}[ht]
	\footnotesize{\fontfamily{qcr}\selectfont 
		user@HVM-domU:\~\# ./test \\
		Can run this program.  \\
		user@HVM-domU:\~\#  cp test newfile\\
		user@HVM-domU:\~\# ./newfile \\
		-bash: ./newfile: \textcolor{red}{Bad address} \\
		user@HVM-domU:\~\# }
	\caption{Confirmation of program white-listing.}
	\label{fig:prog_run}
\end{figure}





\subsection{Quantification of Processing Overhead}

To quantify the processing overhead incurred by \codeft{ferify} upon an authorized user, we have selected three of the most occurring system calls it traps -- namely, \codeft{open()} (for read \& write), \codeft{rename()} (for moving files), and \codeft{unlink()} (for deleting files) -- and benchmarked their performance in each of these three scenarios:

\begin{enumerate}
    \item[{\bf S1}] \codeft{ferify} is not deployed, i.e., no system call is actually trapped.
    
    \item[{\bf S2}] \codeft{ferify} is deployed, but with an empty SACL; in this case, there is no need to search the SACL for a specific file. 
    
    \item[{\bf S3}] \codeft{ferify} is deployed with a full SACL, i.e., with an entry for each file in the VM. The SACL contains more than 200,000 entries.  
    
\end{enumerate}

\noindent
In addition, we examine whether it is possible to increase the performance of \codeft{ferify} by adjusting its scheduling priority. We consider three cases: ({\bf i}) no adjustment; ({\bf ii}) the processing priority of \codeft{ferify} is maximized using the \codeft{nice} command; and ({\bf iii}) both the processing and I/O priorities are maximized using the \codeft{nice} and \codeft{ionice} commands, respectively. 

The bench-marking of each system call is repeated 20 times for each scenario and scheduling priority combination. Their \emph{average} processing times (in ms) are reported in Table~\ref{tbl:measure1}. 

First, we observe that \codeft{ferify}'s overall processing overhead per system call is in the millisecond range, which is usable for most applications, while the overhead is significant as the processing times jumped by more than one order of magnitude with \codeft{ferify}'s deployment. Second, somewhat surprisingly, we observe that adjusting scheduling priorities had little effect on the processing times. Lastly and importantly, we see that there is \emph{little change of performance} from an empty to full SACL, which indicates that the SACL look-up and permission checking actions incurred a very small portion of the overall overhead. In other words, the time spent by other actions, mainly performed by the core DRAKVUF/LibVMI code for trapping the system calls, might dominate the \codeft{ferify} processing overhead.



\begin{table}[ht]
	\centering
	\caption{\codeft{ferify} processing overhead measurements}
	\label{tbl:measure1}
	\begin{tabular}{c|c|c|c}
		\hline
					&  						  & \multicolumn{2}{c}{Ratio of increase}	   \\
					& Without \codeft{ferify} & With \codeft{ferify} & With \codeft{ferify}\\
		System call & Avg time (msec)		  & empty SACL			 & full SACL 		   \\	

		\hline
		\multicolumn{4}{c}{No scheduler priority set}\\
		\hline
		\codeft{open()} 	& \textbf{0.273} & 5.167  & \textbf{6.385}\\
		\codeft{rename()} 	& \textbf{0.119} & 9.891  & \textbf{15.326}\\
		\codeft{unlink()} 	& \textbf{0.110} & 11.461 & \textbf{14.732}\\
		\hline
		\multicolumn{4}{c}{With best \codeft{nice} value}\\
		\hline
		\codeft{open()} 	& \textbf{0.273} & 6.623  & \textbf{6.389}\\
		\codeft{rename()} 	& \textbf{0.119} & 12.686 & \textbf{14.928}\\
		\codeft{unlink()} 	& \textbf{0.110} & 14.643 & \textbf{14.426}\\	
		\hline
		\multicolumn{4}{c}{With best \codeft{nice} and \codeft{ionice} values}\\
		\hline
		\codeft{open()} 	& \textbf{0.273} & 6.461  & \textbf{7.713}\\
		\codeft{rename()} 	& \textbf{0.119} & 13.000 & \textbf{18.238}\\
		\codeft{unlink()} 	& \textbf{0.110} & 14.926 & \textbf{17.384}\\	
		\hline
	\end{tabular}	
\end{table}

\begin{figure}[!htb]
	\centering
	\includegraphics[width=90mm,scale=1]{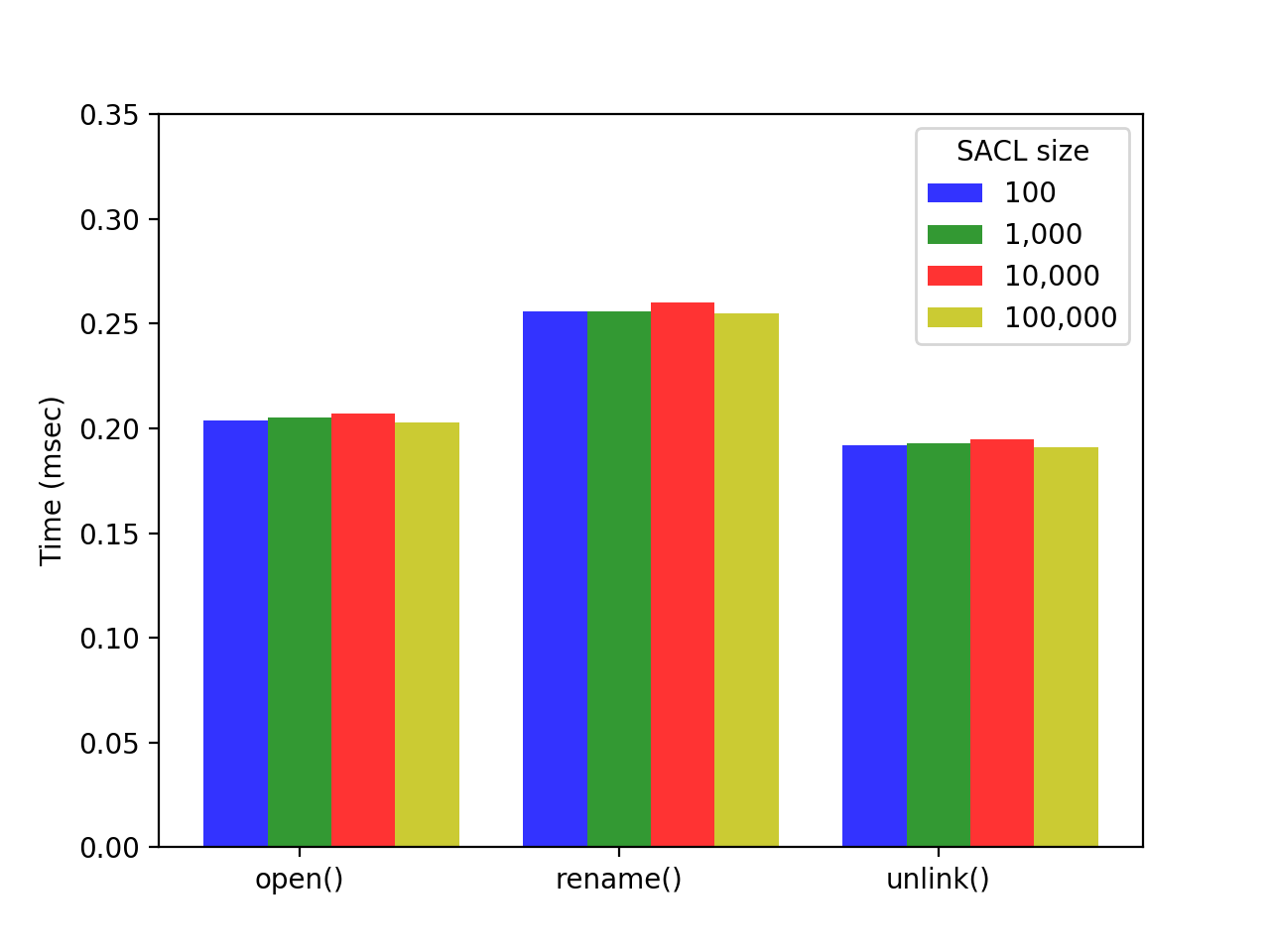}
	\caption{Little changes in ave.\ system call processing times for 4 SACL sizes.}
	\label{fig:measure}
\end{figure}

To further explore this hypothesis, we have conducted additional experiments with \codeft{ferify} deployed using default scheduling priorities, while varying the SACL size four times, to 100, 1000, 10000, and 100000 entries, respectively.  The results are plotted in Fig.~\ref{fig:measure}.  We observe that the overall processing times for each of the three systems call did not change much as the SACL size increases. This is not surprising given that the SACL has been implemented as a hashtable and confirms that SACL processing incurs a very small portion of the overall overhead.

\section{Understanding Overhead of VM Introspection}\label{sec:overhead}

DRAKVUF and LibVMI software hides almost all the system call trapping details from a plugin. To explain why \codeft{ferify} introduces significant processing overhead as presented in Table~\ref{tbl:measure1}, we have to dig deep into the DRAKVUF and LibVMI source code.
After a careful analysis, we discover that the primary contributor to the processing overhead are the multiple CPU context-switches between the hypervisor and the guest VM. The details of our analysis are presented in this section.

\subsection{The Tale of Four Context-Switches} \label{sec:context-switches}

DRAKVUF traps a system call in the same way as a debugger like \codeft{gdb} does. Specifically, DRAKVUF 
replaces the instruction stored in the memory location indicated, with the software breakpoint signal (i.e., opcode \codeft{0xCC}).
Therefore, when the instruction pointer (I.P.) of the VM reaches the trapped address, the CPU will raise a software interrupt. This event, by the way in which Xen handles such interrupts, causes a context switch from the VM to the hypervisor. This context switch is commonly called ``VM-exit'' in the Xen literature. 

Right after this context switch, Xen's master interrupt handler checks for a registered process in Dom0 to handle the interrupt. If no such process exists the hypervisor propagates the interrupt to the VM, to be handled by the guest OS. 
Therefore when \codeft{ferify} is running and 
registered for the interrupt, the master handler will select it to respond to the interrupt.

As soon as \codeft{ferify} completes its checking of file access permissions and clears the relevant register(s) when the decision is to deny the access, DRAKVUF must, again like a debugger, replace the breakpoint signal with the original instruction to allow the system call to proceed. 
Furthermore, DRAKVUF must also inject the breakpoint signal back \emph{after execution of only one instruction} in order to reset the trap and \emph{not miss} any future instance of the system call.  
Currently, there is only one known way to meet the stringent timing requirement for resetting the trap; that is, to put the original process in the single-step execution mode. DRAKVUF accomplishes this by setting a CPU register (flag), just as a debugger would do, before resuming the trapped process. Note that another context switch happens as a result, commonly called ``VM-entry''. Then, after execution of the first instruction, the CPU will raise an exception due to the single-stepping mode. This will be caught by the hypervisor and its interrupt handler, causing a second VM-exit, as illustrated in the figure below. At this point DRAKVUF knows that one instruction has been executed and it is safe to re-inject the breakpoint. After doing so, it clears the CPU flag for single-stepping and returns the trapped process to the normal execution mode, which causes a second VM-entry. 

\begin{figure}[!htb]
\centering
\small

\begin{tikzpicture}

\node[anchor= north west, label={[yshift=-0.1cm]north:System Call Code Block \tab\tab\tab Event}] at (3, 2) {};

\draw[anchor= north west, draw=black, -](0,2) -- (7.5, 2);
\draw[anchor= north west, draw=black, -](0,0) -- (7.5, 0);
\draw[anchor= north west, draw=black, -](0,-2.3) -- (7.5, -2.3);
\draw[anchor= north west, draw=black, -](0,-4.8) -- (7.5, -4.8);
\draw[anchor= north west, draw=black, -](0,-6.5) -- (7.5, -6.5);
\draw[anchor= north west, draw=black, -](0,-8.6) -- (7.5, -8.6);


\node[anchor= north west, label={[yshift=-0.1cm]north:\begin{tabular}{cc}
	\tab0x...: \textcolor{blue}{$I_1~~I_2$ ..\tab\tab} &   
	The first instruction of \\
	& system call (\textcolor{blue}{$I_1$}) is replaced\\
	& by software breakpoint. \\  
    & This needs to happen \\
	& to trap the system call \\
	\end{tabular}}] at (3.5, 0) {};

\draw[anchor= north west, draw=black, - ](1.50, 1.0) -- (1.25, 1.80);
\node[anchor= north west, rectangle, minimum width=5mm, minimum height=4mm, label={[yshift=-0.4cm]\textcolor{blue}{0xCC}}] at (1.13, 1) {};


\node[anchor= north west, label={[yshift=-0.1cm]north:\begin{tabular}{cc}
                \tab0x...: \textcolor{blue}{CC $I_2$ ..\tab\tab} &   
                    System call is just invoked.\\
                    & A hypervisor-VM context \\
                    & switch occurs due to software\\
                    & interrupt.\\
                    & \emph{\textcolor{red}{VM-exit}}~~\#1
                    \end{tabular}}] at (3.5, -2) {};

\draw[anchor= north west, draw=black, ->](1.2,-0.95) -- (1.2, -0.55);
\node[anchor= north west, rectangle, draw=black, minimum width=5mm, minimum height=4mm] at (0.95, -0.1) {};
\node[anchor= north west, rectangle, minimum width=5mm, minimum height=4mm, label={[yshift=-0.4cm]{\bf I.P}}] at (0.95, -0.95) {};


\node[anchor= north west, label={[yshift=-0.1cm]north:\begin{tabular}{cc}
	\tab0x...: \textcolor{blue}{$I_1~~I_2$ ..\tab\tab} &   
	The breakpoint is replaced\\
	& by the original instruction\\
	& in order to resume VM \\
	& execution after  \\
	& single-step is enabled.\\
	& \emph{\textcolor{red}{VM-entry}}~~\#1
	\end{tabular}}] at (3.5, -4.65) {};

\draw[anchor= north west, draw=black, ->](1.35,-3.3) -- (1.35, -2.9);
\node[anchor= north west, rectangle, draw=black, minimum width=5mm, minimum height=4mm] at (1.1, -2.45) {};
\node[anchor= north west, rectangle, minimum width=5mm, minimum height=4mm, label={[yshift=-0.4cm]{\bf I.P}}] at (1.1, -3.3) {};


\node[anchor= north west, label={[yshift=-0.1cm]north:\begin{tabular}{cc}
	\tab0x...: \textcolor{blue}{$I_1~~I_2$ ..\tab\tab} &   
	The execution of \\
	& ~\textcolor{blue}{$I_1$} causes a new trap~~ \\
	& due to single-step mode.\\
	& \emph{\textcolor{red}{VM-exit}}~~\#2
	\end{tabular}}] at (3.5, -6.4) {};

\draw[anchor= north west, draw=black, ->](2.05,-5.75) -- (2.05, -5.35);
\node[anchor= north west, rectangle, draw=black, minimum width=4.5mm, minimum height=4mm] at (1.8, -4.9) {};
\node[anchor= north west, rectangle, minimum width=5mm, minimum height=4mm, label={[yshift=-0.4cm]{\bf I.P}}] at (1.8, -5.75) {};


\node[anchor= north west, label={[yshift=-0.1cm]north:\begin{tabular}{cc}
	\tab0x...: \textcolor{blue}{CC~~$I_2$ ..\tab\tab} &   
	  The first instruction is \\
	& replaced by interrupt \\
	& and single-step is disabled \\
	& before resuming VM.\\
	& \emph{\textcolor{red}{VM-entry}}~~\#2
	\end{tabular}}] at (3.5, -8.5) {};

\draw[anchor= north west, draw=black, ->](2,-7.5) -- (2, -7.1);
\node[anchor= north west, rectangle, draw=black, minimum width=4.5mm, minimum height=4mm] at (1.77, -6.65) {};
\node[anchor= north west, rectangle, minimum width=5mm, minimum height=4mm, label={[yshift=-0.4cm]{\bf I.P}}] at (1.75, -7.5) {};

\end{tikzpicture}
\caption{Illustration of context-switch events for trapping a system call}
\label{fig:libvmi}
\end{figure}

It is well known that a context switch (VM-exit or VM-entry) incurs a significant amount of overhead due to the need for saving the complete state of the VM or hypervisor into memory prior to performing the switch. 
Requiring four context switches for trapping one system call, the current DRAKVUF and LibVMI implementation is unsurprisingly the primary source of \codeft{ferify} processing overhead. More importantly, this overhead is a more general problem, impacting all systems that use DRAKVUF and LibVMI to trap processes running on guest VMs.  Therefore, we have investigated ways to reduce the number of context switches for trapping system calls in a VM, leading to one possible solution, which we will present in the next section.


\subsection{A Proposal for Reducing VMI Overhead}

The question of how to minimize context switches between a guest VM and the hypervisor in order to improve system performance has been studied in other settings~\cite{b21-1,b22-1}. In this section, we propose a solution that can cut the number of context switches from four to two for a hypervisor based VM introspection (VMI) system such as \codeft{ferify} to trap a system call from a VM. 


Specifically, we observe that the second pair of VM-exit and VM-entry context switches, as illustrated in Fig.~\ref{fig:libvmi}, is \emph{necessary only} because the first instruction of the system call code needs be replaced by the software interrupt signal \emph{immediately} after each execution in order to not miss any future invocation of the system call. In other words, the current DRAKVUF/LibVMI implementation requires the first instruction to toggle between two opcodes with a timing requirement that can only be met by putting the calling process into the single step mode.  However, if the first instruction of the system call were the ``do nothing'' (NOP) (i.e., opcode \codeft{0x90} for an Intel x86 CPU), DRAKVUF would be able to replace it with the software interrupt signal \emph{permanently} to trap the system call over and over again, with only two context switches at each time. To avoid an infinite loop, the I.P. of the VM's CPU will be incremented to point to the next valid instruction. 

Therefore, we propose to prepend the code of each trapped system call with the NOP instruction, resulting in one pair of context switches for each trapping, as illustrated in Fig.~\ref{fig:libvmi2}. There are several ways of accomplishing the prepending. One may revise the source code for the object code of the system calls. For a relatively less intrusive approach, we suggest to add such an option to C compilers. 

\begin{figure}[!htb]
	\centering
	\small
	
	\begin{tikzpicture}

	\node[anchor= north west, label={[yshift=-0.1cm]north: System Call Code Block \tab\tab\tab Event}] at (3, 2) {};
	
	\draw[anchor= north west, draw=black, -](0,2) -- (7.5, 2);
	\draw[anchor= north west, draw=black, -](0,0) -- (7.5, 0);
	\draw[anchor= north west, draw=black, -](0,-2.1) -- (7.5, -2.1);
	\draw[anchor= north west, draw=black, -](0,-4.3) -- (7.5, -4.3);

	
	\node[anchor= north west, label={[yshift=-0.1cm]north:\begin{tabular}{cc}
		\tab0x...: \textcolor{blue}{90~~$I_1$ ..\tab\tab} &   
		1st instruction of system \\
		& call (i.e., NOP) is replaced \\
		& by software breakpoint. \\
		& This needs to happen \\
		& to trap the system call.\\
		\end{tabular}}] at (3.5, 0) {};
	
    \draw[anchor= north west, draw=black, -](1.40, 1) -- (1.25, 1.78);
	\node[anchor= north west, rectangle, draw=black, minimum width=5mm, minimum height=4mm] at (1.08, 1.85) {};
	\node[anchor= north west, rectangle, minimum width=5mm, minimum height=4mm, label={[yshift=-0.4cm]0xCC}] at (1.08, 1) {};
	

	\node[anchor= north west, label={[yshift=-0.1cm]north:\begin{tabular}{cc}
		\tab0x...: \textcolor{blue}{CC~~$I_1$ ..\tab\tab} 
& System call is just invoked.\\
& A VM-hypervisor context \\
& switch occurs due to \\
& software interrupt.\\
& \emph{\textcolor{red}{VM-exit}}
		\end{tabular}}] at (3.5, -2) {};
	
	\draw[anchor= north west, draw=black, ->](1.25,-0.95) -- (1.25, -0.55);
	\node[anchor= north west, rectangle, draw=black, minimum width=5mm, minimum height=4mm] at (1.0, -0.15) {};
	\node[anchor= north west, rectangle, minimum width=5mm, minimum height=4mm, label={[yshift=-0.4cm]{\bf I.P}}] at (0.95, -0.95) {};
	
	
	\node[anchor= north west, label={[yshift=-0.1cm]north:\begin{tabular}{cc}
		\tab0x...: \textcolor{blue}{CC~~$I_1$ ..\tab\tab} 
        & A hypervisor-VM context \\
        & switch occurs to resume  \\
		& VM execution. The instruction  \\
		& pointer moves forward.\\
        & \emph{\textcolor{red}{VM-entry}} \\
        & 
	\end{tabular}}] at (3.5, -4.5) {};

\draw[anchor= north west, draw=black, ->](1.63,-3.15) -- (1.63, -2.7);
\node[anchor= north west, rectangle, draw=black, minimum width=5mm, minimum height=4mm] at (1.40, -2.3) {};
\node[anchor= north west, rectangle, minimum width=5mm, minimum height=4mm, label={[yshift=-0.4cm]{\bf I.P}}] at (1.40, -3.15) {};
	
	\end{tikzpicture}
	\caption{Proposed solution incurs only two context switches}
	\label{fig:libvmi2}
\end{figure}


\section{Discussion} \label{sec:discussion}


In this section we discuss possible extensions of \codeft{ferify} and its current limitations.

\subsection{Potential Extensions}

\codeft{ferify} logically should eventually be deployable to desktops, cloud data centers, mobile devices\footnote{LibVMI already supports ARM Cortex-A15 architectures} and IoT systems, due to its relative ease of deployment, for not requiring any kernel modification to the VMs. 
The ability to perform independent access control of individual files makes \codeft{ferify} a unique building block for creating additional VM security solutions. For brevity, we describe two of such extensions as follows. 

First, we envision to extend \codeft{ferify} to generalize the concept of \emph{immutable} file and ``lock down'' parts of a running VM to maximize the availability and integrity of certain system services (e.g., networking) as well as user applications (e.g., database and web severs) during their deployment. In other words, we are interested in elevating the abstraction of protection from files to services and applications. 

Second, we are intrigued by the possibility of using 
\codeft{ferify} to obfuscate an application work-flow to 
further enhance data protection for a VM. For example, a database operator may grant write permissions only to a randomly generated sequence of \codeft{uid}s through \codeft{ferify} and modify the querying process to require forking an ephemeral thread with the correct next \codeft{uid} in that sequence in order to write into the database. This is possible because \codeft{ferify} supports sharing of secrets, in this case a \codeft{uid} sequence, with authorized users or processes in a manner transparent to the VM.

\subsection{Limitations}

While \codeft{ferify} is able to detect and prevent a range of zero-day attacks, the protection is only limited to specific files defined in the SACL.  Also,
the root security mechanism of \codeft{ferify} is dependent on the system calls of the kernel and the system is therefore dependent on the kernel not having malware placed in the kernel during start-up. We decided not to expand further on kernel protection during boot-time; there are already solutions, like vTPM~\cite{b16-1}, which implement a more sophisticated way of securing the system's boot procedure and verifying that the code launched by firmware is trusted. 


Additionally, \codeft{ferify} is currently limited in support of multi-thread processes and multi-core systems. To support multi-threaded applications, more work is necessary to examine the Linux kernel's \codeft{task\_struct} in order to identify where the required thread information is stored and determine how to ensure their integrity. For multi-core systems, although the way how we have utilized DRAKVUF's capabilities should allow direct implementation without further modification, additional testing is required to confirm this is the case.

\section{Conclusion}\label{sec:conclude}

In this research we developed and evaluated \codeft{ferify}, an out-of-guest file protection system capable of detecting and even preventing a range of zero-day attacks against a specific VM. By enforcing totally independent file access control policy, supporting hypervisor based user authentication, and requiring no footprint in the VM, the system has unique advantages over existing security solutions based on known attack signatures. In addition, we performed an investigation into the observed high processing overhead from trapping of system calls, one of key features enabled by current VMI introspection software and used by \codeft{ferify}, which led to a general solution that could potentially cut that overhead by half. Finally, we observe that \codeft{ferify} only scratches the surface of what is possible of leveraging the hypervisor to achieve fine grain user data protection on an VM, and the topic is increasingly fundamental given the seemingly inevitable technology transition towards cloud and fog computing.

\nocite{b2, b3, b4, b5, b6, b7, b8, b9, b11, b12, b13, b14, b15, b16, b17, b18, b19, b20}







\begin{thebibliography}{00}

\bibitem{b1} A. Baliga, L. Iftode, and X. Chen, ``Automated containment of rootkits attacks,'' Computers \& Security, vol. 27, no. 7, pp. 323--334, 2008.

\bibitem{b2} A. Bauman, M. Peinado, and G. Hunt, ``Shielding applications from an untrusted cloud with Haven,'' ACM Transactions on Computer Systems (TOCS), vol. 33, no. 3, 2015.

\bibitem{b3} E. Bauman, A. Gbadebo, and L. Zhiqiang, ``A survey on hypervisor-based monitoring: Approaches, applications, and evolutions'' ACM Computing Surveys (CSUR), vol. 48, no. 1, 2015.

\bibitem{b4} X. Chen et al. ``Overshadow: a virtualization-based approach to retrofitting protection in commodity operating systems,'' In ACM SIGARCH Computer Architecture News, vol. 36, pp. 2--13, 2008.

\bibitem{b5} M. Crawford, and G. Peterson, ``Insider threat detection using virtual machine introspection,'' In Proc. IEEE Hawaii International Conference on System Sciences, pp. 1821--1830, 2013.

\bibitem{b6} B. Dolan-Gavitt, T. Leek, M. Zhivich, J. Giffin, and W. Lee, ``Virtuoso: Narrowing the semantic gap in virtual machine introspection,'' In Proc. IEEE Symposium on Security and Privacy, pp. 297--312, 2011.

\bibitem{b7} G.W. Dunlap, et al. ``Revirt: Enabling intrusion analysis through virtual machine logging and replay,'' In ACM SIGOPS Operating Systems Review, pp. 211--224, 2002.

\bibitem{b8} T. Garfinkel, and M. Rosenblum, ``A virtual machine introspection based architecture for intrusion detection,'' In Proc. NDSS,  pp. 191--206, 2003.

\bibitem{b9} O.S. Hofman, et al. ``Inktag: Secure applications on an untrusted operating system,'' In ACM SIGARCH Computer Architecture News, pp. 265–-278, 2013.

\bibitem{b10} X. Jiang, X. Wang, and D. Xu, ``Stealthy malware detection through VMM-based out-of-the-box semantic view reconstruction,'' In Proc. ACM CCS, pp. 128–-138, 2007.

\bibitem{b11} T. K. Lengyel, et al. ``Scalability, fidelity and stealth in the DRAKVUF dynamic malware analysis system,'' In Proc. Annual Computer Security Applications Conference, 2014.

\bibitem{b12} P. Macko, M. Chiarini, and M. Seltzer, ``Collecting provenance via the Xen hypervisor,'' In Proc. USENIX Workshop on the Theory and Practice of Provenance, 2011.

\bibitem{b13} M. R. Nasab, ``Security functions for virtual machines via introspection,'' Master’s Thesis, Chalmers University of Technology, 2012

\bibitem{b14} D. Ott, ``Virtualization and Performance: Understanding VM Exits''. Available online: https://software.intel.com/en-us/blogs/2009/06/25/virtualization-and-performance-understanding-vm-exits [Last accessed on October 28, 2017].

\bibitem{b15-1} B.D. Payne, ``Simplifying Virtual Machine Introspection Using LibVMI,'' Sandia Labs Tech.\ Report, SAND2012-7818, September 2012.

\bibitem{b15} B.D. Payne, M. Carbone, M. Sharif, and W. Lee, ``Lares: An architecture for secure active monitoring using virtualization,'' In Proc. IEEE Symposium on Security and Privacy, pp. 233--247, 2008.

\bibitem{b16-1} R. Perez, R. Sailer, and L. van Doorn, "vTPM: virtualizing the trusted platform module," In Proc. 15th Conf. on USENIX Security Symposium, pp. 305-320. 2006.

\bibitem{b16} R. Riley, X. Jiang, and D. Xu, ``Multi-aspect profiling of kernel rootkit behavior,'' In Proc. ACM European conference on Computer systems, pp. 47–-60, 2009.

\bibitem{b17} R. Sailer et al. ''Building a MAC-based security architecture for the Xen open-source hypervisor,'' In Proc. Annual Computer Security Applications Conference, 2005. 

\bibitem{b18} A. Seshardi, M. Luk, N. Qu, and A. Perrig, ``Secvisor: A tiny hypervisor to provide lifetime kernel code integrity for commodity OSes,'' In ACM SIGOPS Operating Systems Review, vol. 41, pp. 335--350, 2007.

\bibitem{b19} M.I. Sharif, W. Lee, W. Cui, and A. Lanzi, ``Secure in-VM monitoring using hardware virtualization,'' In Proc. ACM CCS, pp.~ 477--487, 2009.

\bibitem{b20} D. Srinivasan, Z. Wang, X. Jiang, and D. Xu, ``Process out-grafting: an efficient out-of-VM approach for fine-grained process execution monitoring,'' In Proc. ACM CCS, pp.~363--374, 2011.

\bibitem{b21} A. Srivastana, and J. Giffin, ``Efficient protection of kernel data structures via object partitioning,'' In Proc. Annual Computer Security Applications Conference, pp.~429--438, 2012.

\bibitem{b21-1} X. Wang, et al. ``Detecting and Analyzing VM-exits,'' In Proc. IEEE International Conference on Computer and Information Technology, 2010.

\bibitem{b22} Y.-M. Wang, D. Beck, B. Vo, R. Roussev, and C. Verbowski, ``Detecting stealth software with Strider Ghostbuster,'' In Proc. IEEE International Conference on Dependable Systems and Networks, pp.~368--377, 2005.

\bibitem{b22-1} S. Xi, J. Wilson, C. Lu, and C. Gill, ``RT-Xen: towards real-time hypervisor scheduling in Xen,'' In Proc. ACM International conference on Embedded software, 2011.

\bibitem{b23} X. Xiong, D. Tian, and P. Liu, ``Practical protection of kernel integrity for commodity OS from untrusted extensions,'' In Proc. NDSS, 2011.



\end{thebibliography}
\end{document}